\documentclass{PoS}
\usepackage{psfrag}

\newcommand{\beqin}[1]{$ #1 $}
\newcommand{\bra}[1]{\langle #1|} 
\newcommand{\ket}[1]{|#1\rangle}

\newcommand{\lt}{\left}
\newcommand{\rt}{\right}

\newcommand{\tev}{\,\mbox{TeV}}
\newcommand{\gev}{\,\mbox{GeV}}

\newcommand{\ov}{\overline}

\newcommand{\imag}{{\rm Im}\,}
\newcommand{\real}{{\rm Re}\,}

\newcommand{\fig}[1]{Fig.~\ref{#1}}
\newcommand{\eq}[1]{Eq.~(\ref{#1})}
\newcommand{\eqsand}[2]{Eqs.~(\ref{#1}) and (\ref{#2})}

\title{Direct CP violation in $K\to \pi \pi$ decays and supersymmetry}

\ShortTitle{Direct CP violation in $K\to \pi \pi$}

\author{Teppei Kitahara\\
        Institute for Theoretical Particle Physics (TTP)\\
        and Institute for Nuclear Physics (IKP)\\
        Karlsruhe Institute of Technology (KIT)         \\
        76131 Karlsruhe, Germany\\ 
        E-mail: \email{teppei.kitahara@kit.edu}}

\author{\speaker{Ulrich Nierste}\\
        Institute for Theoretical Particle Physics (TTP)\\
        Karlsruhe Institute of Technology (KIT)         \\
        76131 Karlsruhe, Germany\\ 
        E-mail: \email{ulrich.nierste@kit.edu}}

\author{Paul Tremper\\
        Institute for Theoretical Particle Physics (TTP)\\
        Karlsruhe Institute of Technology (KIT)         \\
        76131 Karlsruhe, Germany\\ 
        E-mail: \email{paul.tremper@kit.edu}}

\abstract{The quantities $\epsilon_K^\prime$ and $\epsilon_K$
        measure the amount of direct and indirect CP violation in $K\to
        \pi\pi$ decays, respectively.  Using the recent lattice results
        from the RBC and UKQCD Collaborations and a new compact
        implementation of the $\Delta S=1$ renormalization group
        evolution we predict
\begin{displaymath}
  \real \frac{\epsilon_{K}'}{\epsilon_{K}} = \left(1.06 \pm  5.07  \right) \times
  10^{-4} 
\end{displaymath}
in the Standard Model. This value is $2.8\,\sigma$ below the
experimental value of 
\begin{displaymath}
  \real \frac{\epsilon_{K}'}{\epsilon_{K}} = 
 \left(16.6 \pm 2.3 \right) \times 10^{-4}.
\end{displaymath}
In generic models of new physics the well-understood $\epsilon_K$
precludes large contributions to $\epsilon_K^\prime$, if the new
contributions enter at loop level.  However, one can resolve the tension
in $\epsilon_{K}'/\epsilon_{K}$ within the Minimal Supersymmetric
Standard Model. To this end two features of supersymmetry are crucial:
First, one can have large isospin-breaking contributions (involving the
strong instead of the weak interaction) which enhance
$\epsilon_K^\prime$.  Second the Majorana nature of gluinos permits a
suppression of the MSSM contribution to $\epsilon_K$, because two box
diagrams interfere destructively.}

\FullConference{38th International Conference on High Energy Physics\\
		3-10 August 2016\\
		Chicago, USA}

\begin{document}

\section{Formalism and Standard-Model prediction}
Flavour-changing neutral current (FCNC) transitions of Kaons are
extremely sensitive to new physics and probe mass scales far above the
reach of current high-$p_T$ experiments. $K\to \pi\pi$ decays give
access to two CP-violating quantities, which are related to FCNC
amplitudes changing strangeness $S$ by one or two units, respectively.
To define these quantities $\epsilon_K^\prime$ and $\epsilon_K$ one
first combines the decay amplitudes \beqin{A(K^0\to \pi^+\pi^-)} and
\beqin{A(K^0\to \pi^0\pi^0)} into \beqin{A_0\equiv A(K^0\to (\pi
  \pi)_{I=0})} and \beqin{A_2\equiv A(K^0\to (\pi \pi)_{I=2})} where
\beqin{I} denotes the {strong isospin}. Indirect CP violation 
(stemming from the $\Delta S=2$ box diagrams) is quantified by 
\begin{equation}
\epsilon_K\equiv \frac{A (K_L\to (\pi \pi)_{I=0})}{
                            A (K_S\to (\pi \pi)_{I=0})} =
   (2.228\pm 0.011)\cdot 10^{-3}\cdot e^{i(0.97\pm0.02)\pi/4}
\end{equation}
and was discovered in 1964 \cite{Christenson:1964fg}. The measure 
of direct CP violation, which originates from the $\Delta S=1$ Kaon decay
amplitude, is\footnote{Accidentally, $\epsilon_K^\prime/\epsilon_K$ is essentially real.} 
\begin{equation}
\epsilon_K^\prime \simeq
\frac{\epsilon_K}{\sqrt{2}} \lt[ 
    \frac{  A(K_L\to (\pi\pi)_{I=2}) }{ A(K_L\to (\pi\pi)_{I=0})}
    - 
    \frac{  A(K_S\to (\pi\pi)_{I=2}) }{ A(K_S\to (\pi\pi)_{I=0})}
    \rt]
= (16.6\pm2.3)\cdot 10^{-4} \cdot \epsilon_K .\label{eq:exp}
\end{equation}
This experimental result was established in 1999 and constituted the
first measurement of direct CP violation in any decay \cite{epspexp}.
Adopting the standard phase convention for the elements of the
Cabibbo-Kobayashi-Maskawa (CKM) matrix, the real parts of the isospin
amplitudes are experimentally determined as 
\begin{equation}
\real A_0 =  \left( 3.3201\pm 0.0018 \right)\times
10^{-7}~\gev,\qquad 
\real A_2 =  \left(1.4787 \pm 0.0031\right) \times 10^{-8}~\gev.
\label{eq:a02}
\end{equation}
The master equation for \beqin{\epsilon_K^\prime/\epsilon_K} (see e.g.\
Ref.~\cite{Buras:2015yba})
reads:
\begin{eqnarray}
  \frac{\epsilon_K^\prime}{\epsilon_K} = \frac{\omega_{+}}{\sqrt{2}
    {|}\epsilon_K^{\textrm{exp}}{|} \real A_0^{\textrm{exp}} }
  \left\{ \frac{\imag A_2 }{\omega_{+}} - \left( 1-
      \hat{\Omega}_{\textrm{eff}} \right) \imag A_0 \right\} .
\label{eq:mas}
\end{eqnarray}
Here \beqin{\omega_{+}\simeq \frac{\real A_2}{\real A_0} =
  (4.53 \pm 0.02)\cdot10^{-2}} is determined from the charged
counterparts of $\real A_{0,2}$ and
\beqin{\hat{\Omega}_{\textrm{eff}} = (14.8\pm 8.0)\cdot 10^{-2}}
quantifies isospin breaking. The quantities
$|\epsilon_K^{\textrm{exp}}|$ and $\real A_0^{\textrm{exp}} $ are
also taken from experiment, as indicated.  

\begin{figure}[t]
\centering
{\psfrag{Z}{$\gamma$}
\includegraphics[height=2.5cm]{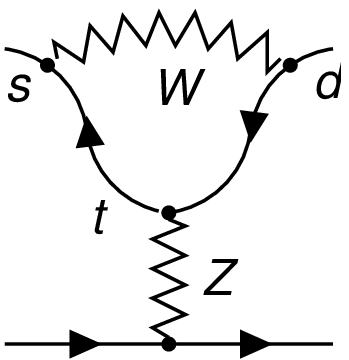}}
\hspace{15mm}
\includegraphics[height=3cm]{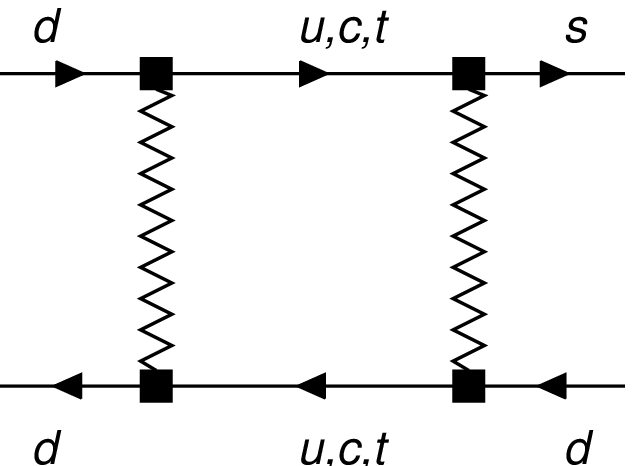}
\caption{Sample diagrams of electroweak penguins and boxes, which 
  contribute to 
  the Wilson coefficient of $Q_8$.\label{fig:diagr}  }
\end{figure}

The important theoretical ingredients encoding potential new-physics
effects are \beqin{\imag A_0} and \beqin{\imag A_2}, which
are calculated from the effective hamiltonian $H^{|\Delta S|=1}$
describing \beqin{s\to d q\ov q} decays. This hamiltonian is known for a
while at the level of next-to-leading-order (NLO) in QCD \cite{nlo} and a 
precise prediction of $\epsilon_K^\prime/\epsilon_K$ is challenged by
the difficulty to calculate the hadronic matrix elements of the 
operators in  $H^{|\Delta S|=1}$. Within the Standard Model (SM) 
\beqin{\imag A_0} is dominated by gluon penguins, with 
roughly 2/3 stemming from the matrix element
\beqin{\bra{(\pi\pi)_{I=0}} Q_6 \ket{K^0}} with the operator
\begin{equation}
Q_6=\ov s_L^j \gamma_\mu d_L^k 
                \sum_q \ov{q}{}_R^k\gamma^\mu q_R^j. 
\end{equation}
About 3/4 of the contribution to \beqin{\imag A_2} 
stems from \beqin{\bra{(\pi\pi)_{I=2}} Q_8 \ket{K^0}} with
\begin{equation}
Q_8=\frac32\ov s_L^j \gamma_\mu d_L^k\sum_q 
        e_q \ov{q}{}_R^k\gamma^\mu q_R^j.
\end{equation}
The Wilson coefficient of $Q_8$ stems from electroweak penguins and box
diagrams (\fig{fig:diagr}).   
Lattice-gauge theory has \beqin{\bra{(\pi\pi)_{I=2}} Q_8
  \ket{K^0}} (and thereby $\imag A_2$) under good control for some
time \cite{Blum:2011ng}, while lattice calculations of 
\beqin{\bra{(\pi\pi)_{I=0}} Q_6 \ket{K^0}} and the other 
matrix elements entering $\imag A_0$ are new \cite{Bai:2015nea}.
The results are consistent with earlier analytic calculations in the 
large-$N_c$ ``dual QCD'' approach \cite{bbg}.
Using these matrix elements from lattice QCD we find \cite{Kitahara:2016nld}
\begin{equation}
  \frac{\epsilon_{K}^\prime}{\epsilon_{K}}  =
  \left( {1.06}\pm  {4.66_{\rm{Lattice}}} \pm {1.91_{\rm{NNLO}}} \pm  
    {0.59_{\rm{IV}}} \pm  { 0.23_{m_t}} \right) \times 10^{-4}.
\label{eq:smnlo}
\end{equation}
The various sources of errors are indicated in the subscripts, with
``NNLO'' referring to unknown higher-orders of the perturbative
expansion and ``IV'' meaning isospin violation.  Adding the errors in
quadrature gives the result in the abstract.  We use the methodology of
\cite{Buras:1993dy}, which exploits the CP-conserving data of
\eq{eq:a02} to constrain the matrix elements. To arrive at \eq{eq:smnlo}
we have implemented a novel compact solution of the renormalization
group equations; the result is in full agreement with the calculation in
Ref.~\cite{Buras:2015yba}. \eq{eq:smnlo} disagrees with the experimental
number in \eq{eq:exp} by 2.8 standard deviations. The original lattice
paper, Ref.~\cite{Bai:2015nea}, quotes a smaller discrepancy. The
discussion at this conference has indicated that the combination of
\eq{eq:a02} with Fierz identities between different matrix elements has
lead to the sharper prediction in
Refs.~\cite{Buras:2015yba,Kitahara:2016nld}.

\section{A supersymmetric solution}
The large factor $1/\omega_{+}$ multiplying $\imag A_2$ in \eq{eq:mas}
renders $\epsilon_K^\prime/\epsilon_K$ especially sensitive to new
physics in the $\Delta I=3/2 $ decay $K\to (\pi\pi)_{I=2}$. This feature
makes $\epsilon_K^\prime/\epsilon_K$ special among all FCNC processes.
However, it is difficult to place a large effect into
$\epsilon_K^\prime$ without overshooting $\epsilon_K$: The SM
contributions to both quantities are governed by the CKM combination
\begin{equation}
  \tau = - \frac{V_{td}V_{ts}^*}{V_{ud}V_{us}^*} \sim (1.5 - i 0.6)\cdot
  10^{-3}. 
\end{equation}
Our quantities scale as
\begin{equation}
\epsilon_K^{\prime\,\rm SM} \propto \imag \frac{\tau}{M_W^2}\qquad \mbox{and}
\qquad
\epsilon_K^{\rm SM} \propto \imag \frac{\tau^2}{M_W^2}.
\label{eq:scsm}
\end{equation}
In new-physics scenarios $\tau $ is replaced by some new $\Delta S=1$
parameter $\delta$ and $M_W$ is replaced by some particle mass $M\gg
M_W$. The new-physics contributions scale as 
\begin{equation}
\epsilon_K^{\prime\, \mathrm{NP}} \propto \imag \frac{\delta}{M^2},\qquad 
\mbox{and}
\qquad
\epsilon_K^{\mathrm{NP}} \propto \imag \frac{\delta^2}{M^2}. \label{eq:scnp}
\end{equation}
If new-physics enters through a loop, the only chance to have a
detectable effect in $\epsilon_K^\prime$ is a scenario with $|\delta|\gg
|\tau|$. Using \eqsand{eq:scsm}{eq:scnp} 
the experimental constraint $|\epsilon_K^{\mathrm{NP}}|
\leq |\epsilon_K^{\rm SM}|$ entails
\begin{equation}
 {\left| {\frac{\epsilon^{\prime\,
          \mathrm{NP}}_K}{\epsilon^{\prime\, \mathrm{SM}}_K}
      } \right|} \leq  
  \frac{\left|\epsilon_K^{\prime\,
         \mathrm{NP}}/\epsilon_K^{\prime\,\rm SM} \right|}{
     \left|\epsilon_K^{\mathrm{NP}}/\epsilon_K^{\rm SM} \right|}
  = {\cal O}\left( \frac{\real \tau}{\real \delta} \right).
\label{eq:sens}
\end{equation}
Thus large effects in $\epsilon_K^\prime$ from loop-induced new physics
are seemingly forbidden. Many studies of $\epsilon_K^\prime$ indeed
involve new-physics scenarios with tree-level contributions to
$\epsilon_K^\prime$ \cite{nptree}, in which the requirement $|\delta|\gg
|\tau|$ can be relaxed. 

\begin{figure}[t]
\centering
\includegraphics[height=2cm]{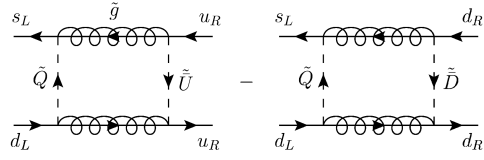}
\caption{``Trojan penguin'' diagrams \cite{Kagan:1999iq}. The difference
of the two boxes contributes to the $\Delta I=3/2$ amplitude and
increases with the mass difference among right-handed up-type ($\tilde U$) and 
down-type ($\tilde D$) squark. $\tilde Q$ denotes a left-handed squark,
which is a strange-down mixture.\label{fig:trojan}}
\end{figure}

\begin{figure}[t]
\centering
\includegraphics[width=0.5\textwidth]{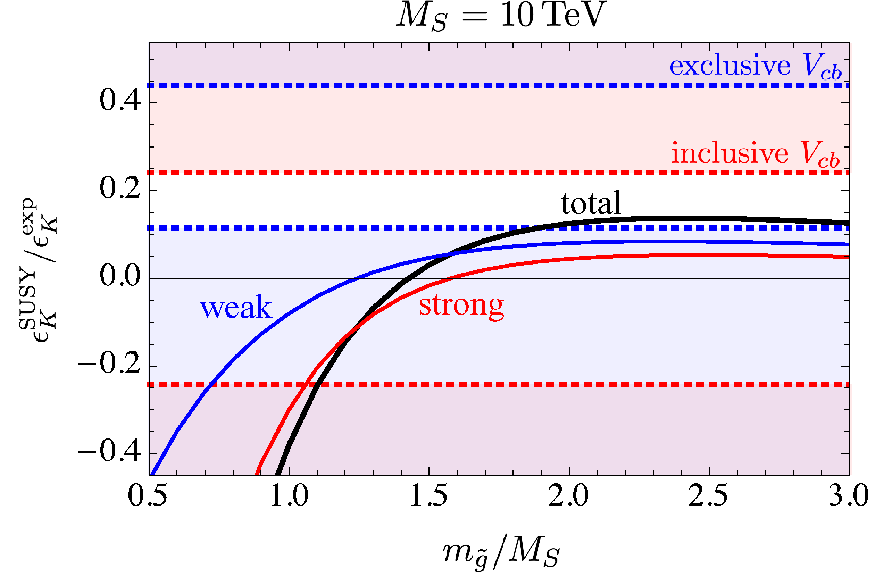} \hfill
\includegraphics[width=0.45\textwidth]{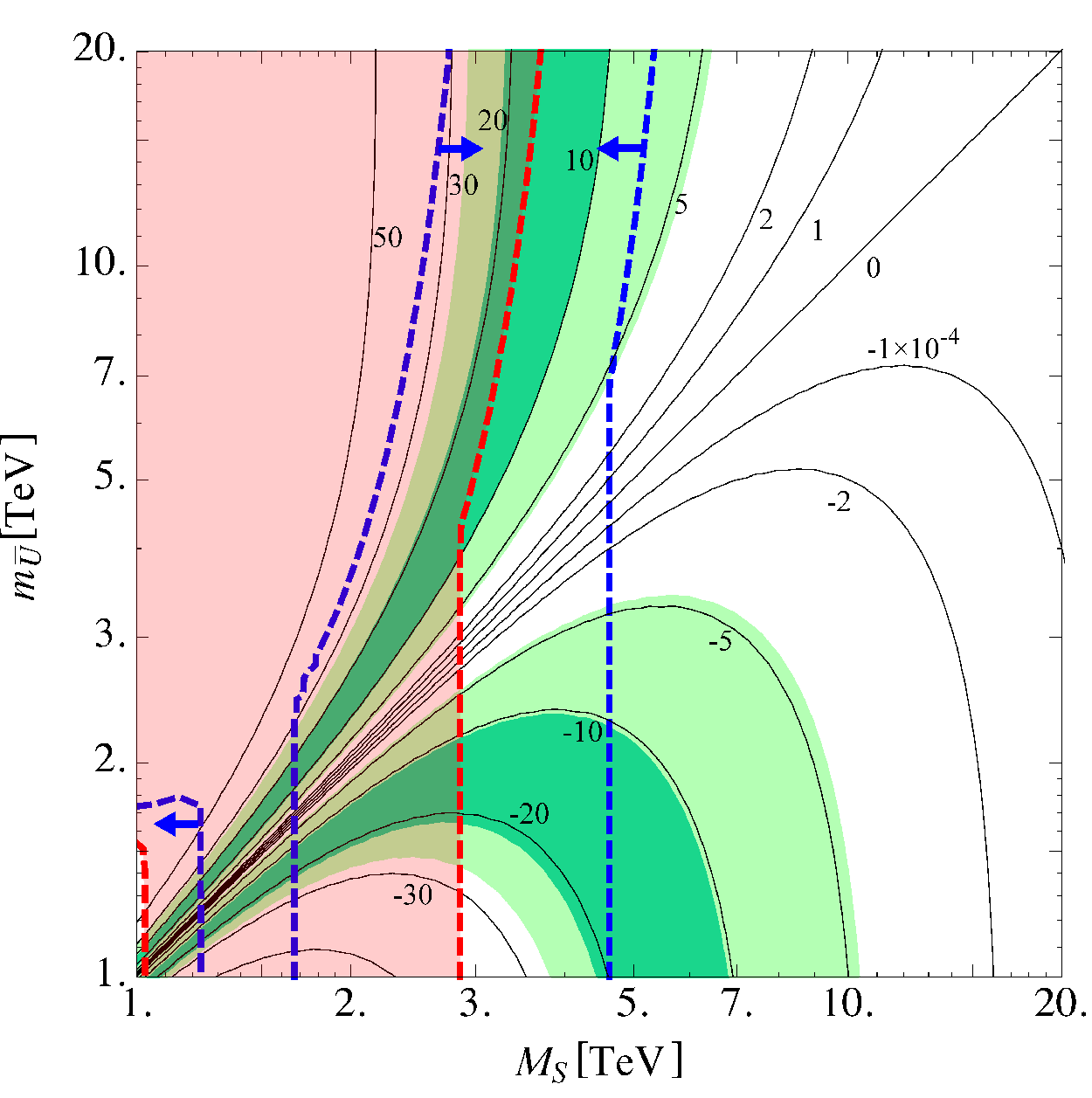}
\caption{Left: $\epsilon_K^{\rm SUSY}/\epsilon_K^{\rm SM}$ as a function of
  $m_{\tilde g}/M_S$ for a common mass $M_S=10\tev$ of all superpartners except 
  the gluino. Right: Parameter region explaining 
  $\epsilon_K^\prime/\epsilon_K$ while complying with the measured
  $\epsilon_K$ for the point $m_{\tilde g}=1.5M_S$ and $M_S=m_{\tilde
    Q}=m_{\tilde D}$. 
  The lines labeled with negative 
  values of the MSSM contribution 
  $\epsilon_K^{\prime\, SUSY}/\epsilon_K$ correspond to correct
  (positive) solutions if the CP phase is appropriately adjusted. 
  The SM prediction
  for $\epsilon_K$ strongly depends on $|V_{cb}|$. The blue (red) 
  lines in both plots delimit the region which complies with 
     $\epsilon_K$ if $|V_{cb}|$ is determined from exclusive (inclusive) 
   $b\to c \ell \nu$ decays. If the exclusive determination is correct, 
   some new physics in $\epsilon_K$ is welcome. In the inclusive case 
   the forbidden region is marked with the red shading. 
   For more details see 
   Ref.~\cite{Kitahara:2016otd}, from which the plots are taken.
   \label{fig:plots}
 }
\end{figure}

Here we present an explanation of the measurement in \eq{eq:exp} by a
supersymmetric loop effect \cite{Kitahara:2016otd}.  We circumvent the
argument in \eq{eq:sens} by exploiting two special features of the
Minimal Supersymmetric Standard Model (MSSM): Firstly, the MSSM permits
large $\Delta I=3/2$ transitions mediated by the strong interaction
(``Trojan penguins'') \cite{Kagan:1999iq}.  These enhanced amplitudes
occur if the mass splitting between the right-handed up and down squarks is
sizable (see \fig{fig:trojan}).
Secondly, the Majorana nature of the gluino permits the suppression of
$\epsilon_K$, which receives contributions from two squark-gluino box diagrams
(``regular'' and ``crossed''). These diagrams cancel each other efficiently, once
the gluino mass $m_{\tilde g}$ and the squark mass $m_{\tilde Q}$ in the loop
satisfy $m_{\tilde g}\geq 1.5 m_{\tilde Q}$ \cite{Crivellin:2010ys}. In our
scenario, the mass scale $M_S$ of the supersymmetric particles is large, of order
$3$--$7\tev$. Squark flavour mixing appears only among the left-handed
doublets. We choose the CP-violating phase  of the (2,1)
element $\Delta^{LL}_{sd}$ of the left-handed squark mass matrix equal to
$\arg(\Delta^{LL}_{sd})=\pi/4$. The results are shown in \fig{fig:plots}

\section{Summary}
Novel lattice results reveal a tension between the measured value of
$\epsilon_K^\prime$ in \eq{eq:exp} and the SM prediction in \eq{eq:smnlo}. Within
the MSSM one can simultaneously enhance $\epsilon_K^\prime$ and suppress unwanted
effects in $\epsilon_K$. Our MSSM scenario works with large superpartner masses in
the 3--7\tev\ range and thereby comply with bounds from collider searches. Crucial
elements are $m_{\tilde g}\geq 1.5 M_{\tilde Q}$, a sizable mass splitting between
right-handed up and down squarks, and flavour mixing among left-handed squarks.

\section*{Acknowledgements}
I thank the organizers for quickly letting me give this talk in a vacant
slot of the timetable.  The presented work has profited from discussions
with Andrzej Buras, Chris Sachrajda, Christoph Bobeth, Martin Gorbahn,
and Sebastian J\"ager.  The work of UN is supported by BMBF under grant
no.~05H15VKKB1.

\end{document}